\documentclass[aps,pre,twocolumn,groupedaddress]{revtex4}
\usepackage{graphicx}

\newcommand \be{\begin{eqnarray}}
\newcommand \ee{\end{eqnarray}}
\newcommand \ba{\begin{eqnarray}}
\newcommand \ea{\end{eqnarray}}
\def\nn{\nonumber}

\begin{document}
\title{Dependence of the transition from Townsend to glow discharge  
on secondary emission}
\author{Yu.P. Raizer$^1$, Ute Ebert$^{2,3}$ and D.D. \v{S}ija\v{c}i\'c$^2$}
\affiliation{$^1$Inst. for Problems in Mechanics of the Russian Academy
of Sciences, Moscow, Russia,}
\affiliation{$^2$CWI, P.O.Box 94079, 1090 GB Amsterdam, The Netherlands,}
\affiliation{$^3$Dept.\ Physics, Eindhoven Univ.\ Techn., The Netherlands.}

\date{\today}

\begin{abstract}
In a recent paper \v{S}ija\v{c}i\'c and Ebert have systematically studied 
the transition from Townsend to glow discharge, refering to older
work from von Engel (1934) up to Raizer (1991), and they stated a strong
dependendence on secondary emission $\gamma$ from the cathode.
We here show that the earlier results of von Engel and Raizer
on the small current expansion about the Townsend limit
actually are the limit of small $\gamma$ of the new expression;
and that for larger $\gamma$ the old and the new results vary 
by no more than a factor of 2. We discuss the $\gamma$-dependence
of the transition which is rather strong for short gaps.
\end{abstract}

\maketitle

In the recent article \cite{DanaStat}, the transition from Townsend 
to glow discharge was re-investigated with analytical and numerical means.
On the analytical side, a systematic small current
expansion about the Townsend limit was performed and it was stated:

``The result agrees qualitatively with the one given by Raizer 
\cite{Raizer} 
and Engel and Steenbeck \cite{Engel}. In particular, the leading
order correction is also of order $\alpha''(j/\mu)^2$.
However, the explicit coefficient of $j^2$ differs:
while the coefficient in \cite{Engel,Raizer} does not
depend on $\gamma$ at all, we find that the dependence on $\gamma$ 
is essential, as the plot of $F$ in Fig.\ 1 (of \cite{DanaStat})
clearly indicates.
In fact, within the relevant range of $10^{-6}\le \gamma\le 10^0$, 
this coefficient varies by almost four orders of magnitude. 
We remark that it indeed would be quite a surprising mathematical result
if the Townsend limit itself would depend on $\gamma$, 
but the small current expansion about it would not.''

Here we remark that while the new systematic calculation
was correct, the interpretation and comparison to earlier work 
requires some correction.

To be precise, the model treated in \cite{Engel,Raizer,DanaStat} and
by many other authors is a one-dimensional time independent Townsend
or glow discharge characterized by the classical equations
\ba
\label{1}
&&\partial_x J_e=|J_e|\;\bar\alpha(|E|),
~~~
\partial_x J_+=|J_e|\;\bar\alpha(|E|),
\\
\label{3}
\label{Poisson}
&&\partial_xE=\frac{\rm e}{\epsilon_0}(n_+-n_e),
\ea
for electron and ion particle current 
\be
\label{4}
J_e = - n_e\mu_eE,~~~J_+=n_+\mu_+E,
\ee 
and electric field $E$. Impact ionization
in the bulk of the discharge is given by the Townsend approximation
\be 
\bar\alpha(|E|)=\alpha_0\;e^{-E_0/|E|}.
\ee
(In \cite{DanaStat}, the generalized case
$\bar\alpha(|E|)=\alpha_0\;\exp\left(-E_0/|E|\right)^s$ was treated.)
Since ions are generated by impact within the gap and drift towards
the cathode, there are no ions at the anode at $x=0$
\be
\label{x0}
n_+(0)=0.
\ee
The ions impacting on the cathode liberate free electrons with rate 
$\gamma$, therefore secondary emission from the cathode is given by
\be
|J_e(d)|=\gamma\;|J_+(d)|,
\ee
where $x=d$ is the position at the cathode.
The electrical potential difference between the electrodes is
\be
U=\int_0^d dx\;E(x),
\ee
and the total electric current is
\be
J={\rm e}\left(n_+\mu_++n_e\mu_e\right)E,~~~\partial_xJ=0.
\ee

It is useful to introduce dimensionless voltage and current
\be 
\label{scale}
u=\frac{U}{E_0/\alpha_0},~~~
\bar j=\frac{J}{\epsilon_0\alpha_0E_0\;\mu_+E_0},
\ee
where $\bar j=j/\mu$ with the definition of $j$ from \cite{DanaStat}.
It should be noted that only bulk gas parameters have been used 
as units; therefore the dimensionless $u$ and $\bar j$ are independent 
of $\gamma$.
 
Further dimensional analysis yields that the current-voltage-characteristics
$u=u(\bar j)$ can depend on three parameters only, namely 
on the dimensionless gap length $L=\alpha_0d$, on the coefficient 
$\gamma$ of secondary emission and on the mobility ratio $\mu=\mu_+/\mu_e$.
In practice, the dependence on the small parameter $\mu$ is almost 
negligibly weak \cite{DanaStat}, therefore $u=u(\bar j,L,\gamma)$.
Here the dimensionless gap length $L$
is related to $pd$ through $L=A\:pd$ as long as the coefficient
$\alpha_0$ is related to pressure like $\alpha_0=Ap$.

How strongly does the characteristics $u=u(\bar j,L,\gamma)$
depend on $\gamma$?
In \cite{DanaStat}, \v{S}ija\v{c}i\'c and Ebert calculated the whole 
Townsend-to-glow regime numerically and derived by expanding 
systematically in powers of current $\bar j$ about the Townsend limit
\ba
\label{exp}
u&=&u_T-A_{SE}\;\bar j^2\;+O(\bar j^3),\\
\label{ASE}
&&A_{SE}=\frac{{\cal E}_T\;\alpha''}{2\;\alpha'}\;
\frac{F(\gamma,\mu)}{(\alpha {\cal E}_T)^3},
\\
&&\alpha({\cal E}_T)=e^{-1/|{\cal E}_T|},
\label{alpha}
\ea
which gave an excellent fit to the numerical solutions. Here
\ba
\label{Fg}
F(\gamma,\mu)&=&\frac{L_\gamma^3}{12}
+(1+\mu)\left(2-L_\gamma-2e^{-L_\gamma}-L_\gamma 
e^{-L_\gamma}\right)
\nonumber\\
&&+\;(1+\mu)^2\left(\frac{1-e^{-2L_\gamma}}{2}
               -\frac{(1-e^{-L_\gamma})^2}{L_\gamma}\right),
\nn\\
L_\gamma&=&\ln\;\frac{1+\gamma}{\gamma},
\ea
and ${\cal E}_T$ and $u_T$ are field and potential in the Townsend limit
of ``vanishing'' current, i.e., with breakdown values 
\be 
\label{ET}
{\cal E}_T=\frac1{\ln (L/L_\gamma)},~~~u_T=\frac{L}{\ln (L/L_\gamma)}.
\ee
The minimal potential $u_T$ is $L_\gamma e^1$, it
is attained for gap length $L=L_\gamma e^1$ on the Paschen curve
$u_T=u_T(L)$ \cite{DanaStat,Engel,Raizer}.

In \cite{DanaStat}, it was argued that the coefficient $A_{SE}$ in
(\ref{exp}) strongly depends on $\gamma$ due to the factor
$F(\gamma,\mu)$ in (\ref{ASE}). This factor strongly depends on $\gamma$,
for small $\gamma$ actually in leading order like $L_\gamma^3/12$.
(Note that there is a discrepancy between equation (50) in
\cite{DanaStat} for $F(\gamma,\mu)$ which is reproduced 
as Eq.~(\ref{Fg}) in the present paper, and the plot in Fig.~1 
of \cite{DanaStat} for $10^{-1}<\gamma<10^0$. Equation (50) in 
\cite{DanaStat} is correct and the figure erroneous.
$F(\gamma,\mu)$ actually varies by five orders of magnitude on
$10^{-6}<\gamma<10^0$, not only by four.)

At this point, the question how the remaining factor
depends on $\gamma$ was omitted. In fact, the denominator 
$(\alpha{\cal E}_T)^3$ in (\ref{exp}) has in leading order 
the same strong dependence on $\gamma$, since 
\be
\frac1{(\alpha{\cal E}_T)^3}=
\left(\frac{L}{L_\gamma}\;\ln(L/L\gamma)\right)^3,
\ee 
according to the Townsend breakdown criterion $\alpha L=L_\gamma$,
cf.\ (\ref{alpha})--(\ref{ET}).
Therefore the leading order dependence on $L_\gamma^3$ of the coefficient
of $\bar j^2$ in (\ref{exp}) is cancelled and replaced by a dependence on 
$L^3$, while the term with $\alpha''$ has the classical explicit form
\be
\frac{{\cal E}_T\;\alpha''}{2\alpha'}=\frac{1- 2 {\cal E}_T}{2 {\cal E}_T}
=\frac{\ln(L/L_\gamma)-2}{2}.
\ee

In \cite{Engel,Raizer}, another small current expansion was derived from
(\ref{1})--(\ref{4}), assuming $n_+\gg n_e$ and $n_+(x)\approx {\rm const.}$
This approximation was criticised in \cite{DanaStat}, since it is in 
contradiction with the boundary condition (\ref{x0}) --- however,
for very small $\gamma$, it is a good approximation in a large part
of the gap. The resulting equations (8.8) and (8.10) from \cite{Raizer} 
read in the notation of the present paper
\ba
\label{UR}
U&=&U_T-\frac{U_T}{48}\; \frac{1-2{\cal E}_T}{2{\cal E}_T}
\left(\frac{J}{J_L}\right)^2,
\\
\label{JL}
&&J_L=\frac{\epsilon_0\mu_+U_T^2}{2 d^3}.
\ea
(Here a misprint in \cite{Raizer} was corrected, namely the missing 
factor $U_T$ in the coefficient of $J^2$ in (\ref{UR}), 
and the factor $1/(8\pi)$ in (8.8) is substituted by $\epsilon_0/2$
in (\ref{JL}), since we here write the Poisson equation (\ref{Poisson}) 
in MKS units rather than in Gaussian units, cf.~(8.6) in \cite{Raizer}.)

In (\ref{UR}), the physical current density $J$ is compared to $J_L$.
$J_L$ is the current density where deviations from the Townsend limit 
through space charges start to occur; it explicitly depends
on $\gamma$ through $U_T$ (\ref{ET}). 

Comparison of the results of \v{S}ija\v{c}i\'c/Ebert (SE) (\ref{exp}) and 
Engel/Raizer (ER) (\ref{UR}) show that the coefficients $A_{SE,ER}$ 
in the expansion (\ref{exp}) are related like 
\be
A_{SE}=A_{ER}\;\frac{12\;F(\gamma,\mu)}{L_\gamma^3}, ~~~
A_{ER}=\frac{1- 2 {\cal E}_T}{2 {\cal E}_T}\;\frac{L^3}{12\;{\cal E}_T^3}.
\ee
The coefficients $A_{SE}$ and $A_{ER}$ depend in the same way on $L$,
and they are essentially independent of $\mu$ for realistic values of $\mu$.
Therefore the ratio $A_{SE}/A_{ER}$ depends only on $\gamma$ as shown 
in Fig.~1. For $\gamma\to0$, the ratio tends to unity. For a large range
of $\gamma$ values, the deviation is not too large, approaching
a factor 0.44 for $\gamma=10^{-1}$.

\begin{figure}[htbp]
  \begin{center}
    \includegraphics[width=0.5\textwidth]{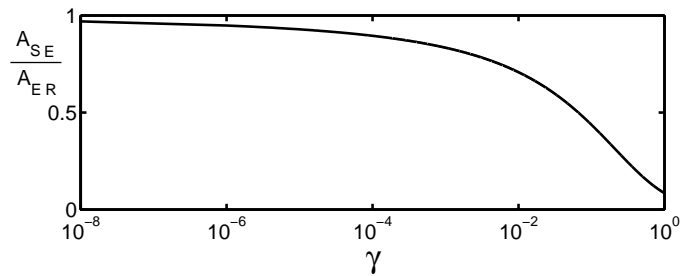} \\
    \caption{The ratio $A_{SE}/A_{ER}$ of the small current expansions
       by \v{S}ija\v{c}i\'c/Ebert and Engel/Raizer as a function of $\gamma$.}
   \label{fig1}
  \end{center}
\end{figure}

 Fig.~2 shows that the factor $A_{SE}$ indeed strongly depends on $\gamma$ for the given L.


\begin{figure}[htbp]
  \begin{center}
    \includegraphics[width=0.49\textwidth]{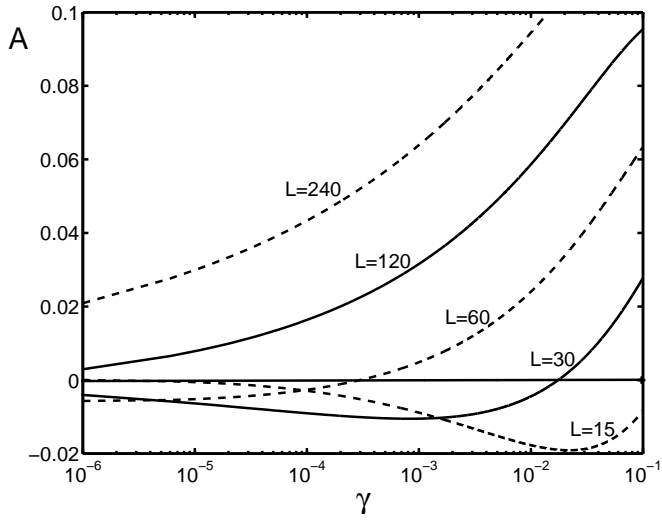} \\
    \caption{The normalized coefficient $A= 24\; A_{SE}/(L^3\ln^4L)$ 
    as a function of $\gamma$ for gap lengths $L=A\:pd=15$, 30, 60, 120, 240 (dashed and solid  lines with labels).}
   \label{fig2}
  \end{center}
\end{figure}

The strong dependence of $A_{SE}$ or $A_{ER}$ on $\gamma$ for a given 
short gap length $L$ means that we can obtain both negative and 
positive differential resistance $dU/dJ$ close to the Townsend limit 
for the same gap length.
Therefore the choice of $\gamma$ is important since it can change
the differential conductivity and therefore the stability of 
a Townsend discharge in a short gap.

\begin{acknowledgments}
Y.R. acknowledges hospitality of CWI Amsterdam 
and D.S. a Ph.D. grant of the Dutch physics funding agency FOM.
\end{acknowledgments}


\end{document}